\documentclass[twocolumn,showpacs,aps,prl]{revtex4}

\usepackage{graphicx}
\usepackage{dcolumn}
\usepackage{amsmath}

\def\PRL{Phys. Rev. Lett. }
\def\PRC{Phys. Rev. C }
\def\PRD{Phys. Rev. D }
\def\PLB{Phys. Lett. B }

\def\etal{\emph{et al.}}

\def\pT{\mbox{$p_T$}}
\def\sqrtsNN{\mbox{$\sqrt{s_{NN}}$}}

\begin{document}

\title{
Azimuthal Anisotropy and Correlations 
in the Hard Scattering Regime at RHIC
}
\author{
C.~Adler$^{11}$, Z.~Ahammed$^{23}$, C.~Allgower$^{12}$, J.~Amonett$^{14}$,
B.D.~Anderson$^{14}$, M.~Anderson$^5$, G.S.~Averichev$^{9}$, 
J.~Balewski$^{12}$, O.~Barannikova$^{9,23}$, L.S.~Barnby$^{14}$
% ...
\\
(STAR Collaboration)
}
% ...
\affiliation{$^5$University of California, Davis, California 95616}
% ...
\affiliation{$^{9}$Laboratory for High Energy (JINR), Dubna, Russia}
% ...
\affiliation{$^{11}$University of Frankfurt, Frankfurt, Germany}
\affiliation{$^{12}$Indiana University, Bloomington, Indiana 47408}
% ...
\affiliation{$^{14}$Kent State University, Kent, Ohio 44242}
% ...
\affiliation{$^{23}$Purdue University, West Lafayette, Indiana 47907}
% ...

\date{\today}

\begin{abstract}
Azimuthal anisotropy ($v_2$) and two-particle
angular correlations 
of high \pT\  charged hadrons have been measured
in Au+Au collisions at $\sqrt{s_{NN}}$=130 GeV for transverse momenta up to 
6 GeV/c, where hard processes are expected to contribute significantly.
The two-particle angular correlations exhibit elliptic flow and
a structure suggestive of fragmentation of high $p_T$ partons.
The monotonic rise of $v_2(p_T)$ for $p_T<2$
GeV/c is consistent with collective hydrodynamical flow calculations.
At $p_T>3$ GeV/c a saturation of $v_2$ is observed
which persists up to $p_T=6$ GeV/c. 

\end{abstract}

\pacs{25.75.Ld}
               
\maketitle

Collisions of heavy nuclei at ultra-relativistic energies
exhibit strong collective flow effects indicative
of a volume of hot matter so dense that descriptions involving hydrodynamic
behavior in a locally thermalized system may apply \cite{v2charged}. 
The azimuthal anisotropy of final state hadrons in 
non-central collisions \cite{olli} 
is sensitive to the system evolution at early
times \cite{sorge}. 
At high $\pT$, a hydrodynamic description of the system may break down as processes involving hard scattering of the initial-state partons are expected to play the dominant role.

Calculations based on perturbative QCD predict that high energy 
partons traversing nuclear matter lose energy through induced gluon radiation 
\cite{energyloss}, where the magnitude of the energy loss is dependent upon
the density of the medium \cite{energyloss2}.
Recent measurements of inclusive charged hadron
distributions in Au+Au collisions at \sqrtsNN=130 GeV
show a suppression of hadron yields at high \pT\  
in central collisions relative to
peripheral collisions and scaled nucleon-nucleon interactions,
consistent with the picture of partonic energy loss in a dense system
\cite{phenix,StarHighpt}.
 The
fragmentation products of partons that have propagated 
through the azimuthally 
asymmetric
system generated by non-central collisions may exhibit azimuthal
anisotropy due to energy loss and the azimuthal dependence of the path
length, providing important information about the initial
conditions and dynamics in a heavy ion collision
\cite{wang,glv}. 

The azimuthal anisotropy of an event in momentum space 
is quantified by the coefficients of the  Fourier decomposition of the 
azimuthal particle distribution with respect to the reaction plane, with 
the second harmonic coefficient $v_2$ 
referred to as elliptic flow.
These coefficients can be inferred from the particle distribution with respect 
to the estimated reaction plane orientation, corrected for the reaction plane 
resolution, or from two-particle correlation analysis~\cite{voloshin}.  
The methods are identical  
if the azimuthal correlation between particles results solely from their
correlation with the reaction plane.
Correlations
that are localized in both rapidity and azimuthal angle are characteristic
of high energy partons fragmenting into jets of hadrons.
Such short-range correlations may be isolated from elliptic flow using 
two-particle correlation analyses performed in different regions of relative 
pseudorapidity.

The transverse momentum dependence of $v_2$ has been previously measured in 
Au+Au collisions at $\sqrt{s_{NN}}=130$ GeV for 
charged \cite{v2charged} and identified \cite{v2identified} 
particles 
in the region of $p_T<2$ GeV/c. 
Elliptic flow at RHIC can be described by a hydrodynamical 
model for $p_T$ up to 2 GeV/c. 
In this Letter we report the first results on $v_2(p_T)$ of charged particles 
measured in this reaction up to $p_T=6$ GeV/c, together with the analysis 
of two-particle azimuthal correlations among 
high $p_T$ charged particles.

The Solenoidal Tracker at RHIC (STAR) consists of several detector
subsystems in a large solenoidal magnet.  The main tracking detector
is the Time Projection Chamber (TPC), which has wide acceptance in 
pseudorapidity and
complete azimuthal coverage \cite{nimTPC}.  For this analysis, the
full data set from the first year data taking of the STAR experiment was used, 
consisting of 300K
minimum-bias and 400K centrally triggered events.  The minimum-bias
data contain hadronic Au+Au interactions at \sqrtsNN=130 GeV 
corresponding to $\sim$90\% of the
geometric cross section $\sigma_{geo}$, while the centrally
triggered data provide an unbiased
event sample 
for the most central 10\% of the minimum-bias data set. 
The trigger conditions and event and track selection cuts for $v_2$ analysis 
are identical to those used previously~\cite{v2identified}.

The reaction plane analysis method involves the calculation 
of the orientation of the event plane, which is an 
experimental estimator of the true reaction plane angle. 
For this analysis, 
the second harmonic event plane angles, $\Psi_2$, 
were calculated for the full event and 
two subevents, consisting of randomly selected
particles from the same event. 
The results are insensitive to the selection method (random,
pseudorapidity, or charge sign)  
for assigning particles to subevents \cite{v2charged}. 

Jets may bias the reconstruction of the reaction plane 
if the intra-jet correlations produce asymmetries which are of similar
magnitude to that due to collective flow.
Systematic studies were undertaken to 
assess the bias. 
In general, the products of jet fragmentation have higher $p_T$ than
other particles produced in a collision \cite{UA1}. Only low $p_T$ particles 
($p_T<2$ GeV/c) were selected to calculate 
the event plane orientation. Using
$p_T$-cutoffs of 1.5, 1.0 and 0.5 GeV/c resulted in statistically
consistent $v_2$ values.
In addition, two sets of events were analyzed,
the first  
containing only events with a particle of $p_T>3$ GeV/c and the second 
with all events, separately.  
The values of $v_2$ as a function of $p_T$ were the same in both
data sets.
 To ensure that particles produced 
within a jet do not affect the reaction plane reconstruction,
all subevent particles in a pseudorapidity 
region of $|\Delta\eta|<0.5$ around the highest $p_T$ particle
in the event were excluded and the results were insensitive to this procedure. 
Varying the track selection criteria (distance of closest approach to the 
primary vertex, number of measured space points, etc.) also made no 
significant difference.

Finite momentum resolution at high transverse momenta combined with a
rapid decrease of hadron yield with increasing $p_T$ may
cause flattening of the $p_T$-dependence of $v_2$.  The momentum
resolution has been determined by embedding simulated single tracks
into real raw data events. For the cuts used in this analysis, 
the momentum resolution for $p_T>1.5$ GeV/c is parameterized as $\delta
p_T/p_T=0.013 + 0.014p_T$/(GeV/c) for central events at magnetic 
field $B$=0.25 T. 
 We have studied the possible effects of the
momentum resolution on $v_2$ at high transverse momenta using a
Monte-Carlo simulation by generating particles with a power-law
$p_T$-distribution \cite{StarHighpt} and with various $v_2(p_T)$ 
dependencies.  The
estimated relative systematic error on $v_2$ due to momentum
resolution is 5\% at $p_T$=5 GeV/c.

The reaction plane analysis integrates all possible sources of azimuthal 
correlations, including those unrelated to the orientation of the 
reaction plane. 
The non-flow correlations may be due to resonance decays, 
(mini)jets, final state interactions 
(particularly Coulomb effects), momentum conservation, etc.
The strength of non-flow contributions relative to the 
measured azimuthal asymmetry 
was discussed in \cite{v2charged}, which estimated
that 15-20\% of the $v_2$ signal obtained with a reaction plane analysis 
method is due 
to non-flow correlations. 
A four-particle correlation method \cite{4part} for 
flow measurements reduces non-flow sources to a negligible level. 
The centrality averaged values of $v_2$ from four-particle correlations 
are 15\% lower  than those obtained from the reaction
plane analysis \cite{flowprc}. The ratio of $v_2$ from the two methods is
approximately independent of pseudorapidity and 
transverse momentum within $0.1<p_T<4.0$ GeV/c, the range
accessible to the four-particle correlation method with the 
current statistics \cite{flowprc}. Based on these studies, we assign a 
$^{~+5}_{-20}$\% systematic uncertainty to the $v_2$ values presented. This
uncertainty is highly correlated point-to-point at all $p_T$-values.

\begin{figure}[htb]
\resizebox{0.5\textwidth}{!}{\includegraphics{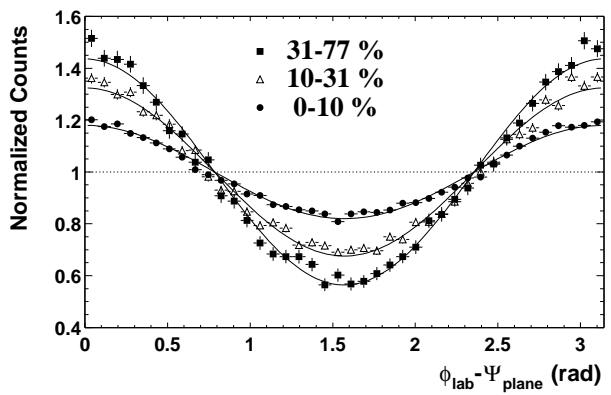}}
  \caption{Azimuthal distributions with respect to the reaction plane of 
charged particles within $2<p_T<6$ GeV/c,  
for three collision centralities.
The percentages are given with respect to the geometrical cross section 
$\sigma_{geo}$.
 Solid lines show fits by $1+2v_2\cos 2(\phi_{lab}-\Psi_{plane})$.}
\label{dndphi}
\vspace{-3.0mm}
\end{figure}

Figure~\ref{dndphi} shows the azimuthal distributions with respect to the reaction plane of charged 
particles within $2<p_T<6$ GeV/c,  
for three collision centralities. The distributions are corrected
for the reaction plane resolution.
All distributions 
exhibit the second harmonic behavior
characteristic of elliptic flow. 
A fit of distributions by $1+2v_2\cos 2(\phi_{lab}-\Psi_{plane})$ 
yields $v_2=0.218\pm 0.003, 0.162\pm 0.002, 0.090 \pm 0.001$ 
for the three centrality bins. 
The errors are statistical only. 
There is large azimuthal anisotropy at high $p_T$ for all centralities.

\begin{figure}[htb]
\resizebox{0.5\textwidth}{!}{\includegraphics{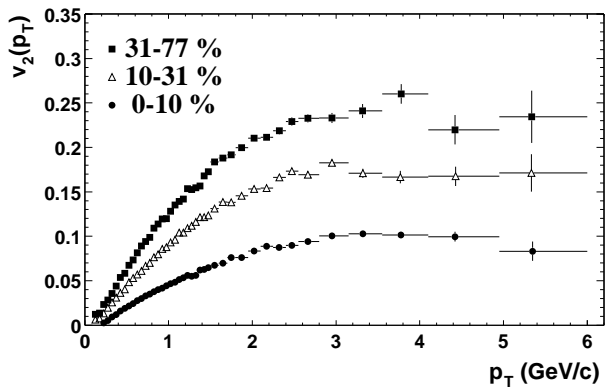}}
  \caption{$v_2(p_T)$ for different collision centralities. 
The errors are statistical only. The systematic uncertainties, which are highly correlated point-to-point, are $^{~+5}_{-20}$\%.}
\label{3cent}
\vspace{-3.0mm}
\end{figure}

Figure~\ref{3cent} 
shows the differential elliptic flow $v_2$ 
\vspace{-3.0mm}
as a function of 
$p_T$ for three collision centralities. The values of $v_2(p_T)$ were 
obtained from the second moment of the distribution of particles with 
respect to the 
reaction plane, i.e. from the average values $\langle \cos 2(\phi_{lab}-\Psi_{plane}) \rangle$, corrected for the reaction plane resolution.
At a given $p_T$, the more
 peripheral collisions have larger $v_2$. 
For all centralities, 
$v_2$ rises linearly up to $p_T=1$ GeV/c, 
then deviates from a linear rise 
and saturates for $p_T>3$ GeV/c. 
The saturation persists up to 6 GeV/c and is in contrast to non-dissipative
hydrodynamical 
calculations, which predict a continuous rise of $v_2$ with increasing
transverse momentum \cite{hydro}. 

\begin{figure}[htb]
\resizebox{0.5\textwidth}{!}{\includegraphics{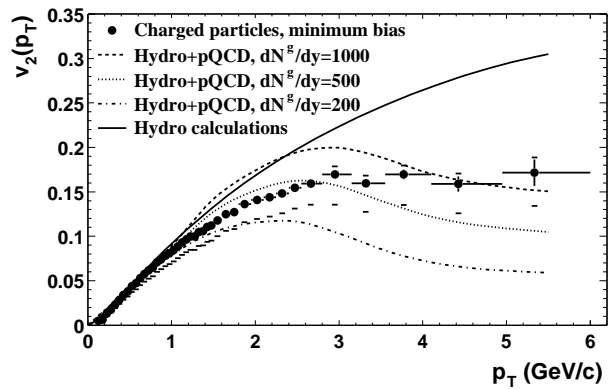}}
  \caption{$v_2(p_T)$ for minimum-bias events (circles).
 The error bars represent the statistical errors and the caps show the 
systematic uncertainty. 
 The data are compared with 
hydro+pQCD calculations \cite{glv} assuming the initial 
gluon density $dN^g/dy=1000$ 
(dashed line), 500 (dotted line), and 200 (dashed-dotted line).
Also shown are  pure 
hydrodynamical calculations \cite{hydro} (solid line). }
\label{minbias}
\vspace{-3.0mm}
\end{figure}

In Ref.~\cite{glv}, 
the particle production is 
decomposed into phenomenological ``soft'' and perturbative QCD calculable 
hard components.
The soft nonperturbative component incorporates hydrodynamic 
elliptic flow, whereas the pQCD calculable part includes energy 
loss (jet quenching). 
In this model the magnitude of $v_2$ at high $p_T$ is 
sensitive to nuclear geometry and the initial gluon density achieved in a collision.
Figure~\ref{minbias} compares the minimum-bias 
differential elliptic flow $v_2(p_T)$ with calculations from Ref.~\cite{glv}.
These calculations also predict a decrease of $v_2$ with increasing $p_T$ at 
high transverse momenta. 
This decrease should even be stronger if transverse 
expansion of the system is taken into account \cite{glvp}. 
A rapid expansion dilutes the initial coordinate 
space azimuthal asymmetry resulting in a reduction of the measured azimuthal 
anisotropies due to energy loss. 
A flavor dependence of both the asymmetry and $p_T$ differential 
particle multiplicities has been suggested as one of 
the possible scenarios for the
 $v_2(p_T)$ behavior \cite{glvp}. 
The observed saturation of $v_2$ at $p_T\sim $2-3 GeV/c can be quantitatively 
reproduced in a parton cascade model
with only elastic rescatterings, but extreme initial gluon 
densities, $dN^g/d\eta\sim 15,000$,  or extreme elastic parton cross sections,
$\sim 45$ mb, are required \cite{molnar}.

In order to verify the existence of a hard scattering and fragmentation 
component at high $p_T$, two-particle angular correlation measurements 
are used \cite{UA1}.  In central heavy-ion collisions, it is impossible
to reconstruct jets fully due to the large overall particle density.
Angular correlations of high $p_T$ particles, however, 
allow for an identification of a hard scattering component 
on a statistical basis.  
The fragmentation of high $p_T$ partons into several particles
results in correlations of hadrons at small $\Delta \eta$, $\Delta \phi$. 
In order to isolate this short-range component of the two-particle 
correlation  function, the azimuthal correlations 
of high $p_T$ particles are measured 
in two regions of relative pseudorapidity.  
At large $\Delta \eta$, we assume 
that the azimuthal correlations are 
free from the fragmentation component.

For the two-particle azimuthal correlation analysis,  
events containing a trigger particle having $4<p_T$(trig)$<6$ GeV/c and $|\eta|<0.7$ 
are used. For these events, we measure the relative azimuthal distribution of
other charged tracks with 2 GeV/c $<p_T<p_T$(trig) and  $|\eta|<0.7$, and the
distribution is normalized to the number of high $p_T$ trigger particles,
\begin{equation} 
\frac{1}{N_{{\rm trigger}}}\frac{dN}{d(\Delta \phi)} \equiv \frac{1}{N_{{\rm trigger}}}\frac{1}{\epsilon} \int d\Delta \eta N(\Delta \phi, \Delta \eta).
\end{equation} 
$N_{{\rm trigger}}$ is the observed number of tracks
satisfying the trigger requirement, and $N(\Delta \phi, \Delta \eta)$
is the number of observed pairs as a function of relative azimuth
($\Delta \phi$) and pseudorapidity ($\Delta \eta$), and $\epsilon$
is the single track efficiency.
Due to the nearly uniform azimuthal acceptance of STAR, no mixed 
event reference is required.
With this definition of the correlation
function, the efficiency for finding the trigger particle cancels, 
and we need only correct 
the data for the efficiency 
of finding the lower $p_T$ particle. This efficiency is determined using
embedding and is 66\% for the tracks used in this analysis.

\begin{figure}[htb]
\resizebox{0.5\textwidth}{!}{\includegraphics{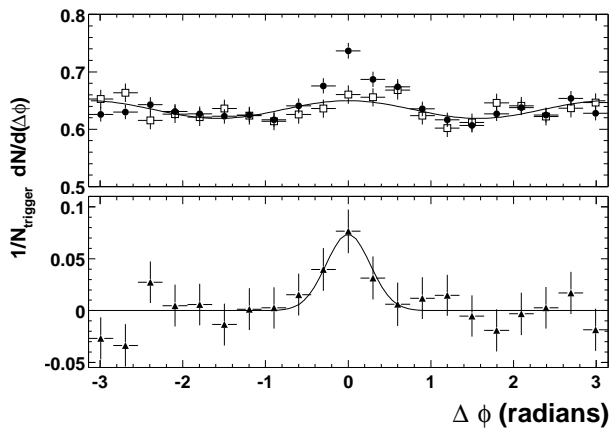}}
\caption{High $p_T$ azimuthal correlation functions for 
central events.  Upper panel: Correlation function 
for $|\Delta \eta|<0.5$ (solid circles) and scaled correlation 
function for $0.5<|\Delta \eta|<1.4$ (open squares). 
Lower panel: Difference of the two correlation functions. 
Also shown are the fits to the data (described in the text).}
\label{c2}
\vspace{-3.0mm}
\end{figure}
Figure \ref{c2} shows the two-particle azimuthal correlation
functions for $|\Delta \eta|<0.5$ and $|\Delta \eta|>0.5$ for central events. 
Here the small relative pseudorapidity correlation function 
is absolutely normalized, while the large relative pseudorapidity 
correlation function has been scaled
to match the small  $|\Delta \eta|$ correlation function in the region 
$0.75 < |\Delta \phi| <2.25$.  
There is an enhancement near $\Delta \phi = 0$. This short range correlation
may
be evidence of hard scattering and fragmentation.  
The large relative
pseudorapidity correlation function is fit by a functional 
form $dN/d(\Delta\phi) \propto 1+2v_2^2\cos(2\Delta\phi)$, expected if the
correlations are due entirely to elliptic flow.  
This fit gives $v_2 = 0.11\pm 0.02$. 

In order to compare this pair-wise method  quantitatively 
to the reaction plane-based $v_2$ analysis shown in Figure~\ref{dndphi}, 
we construct azimuthal correlation functions with a 
trigger particle having $2<p_T<6$ GeV/c. 
For $|\Delta\eta|>0.5$, the azimuthal correlations 
give $v_2$ results consistent with the reaction plane analysis:
$v_2=0.203 \pm 0.012, 0.160 \pm 0.007,$ and $0.091\pm 0.003$ 
for the three centrality bins.
If no pseudorapidity gap is required between two particles, 
the $v_2$ values are 10-15\% higher.

Also shown in Figure \ref{c2} is the difference between the small and large
relative pseudorapidity correlation functions. 
There is an enhancement near zero, and a flat correlation function 
at larger $\Delta \phi$. 
A Gaussian fit gives $\sigma = 0.27\pm0.09$(stat.)$\pm0.04$(sys.) radians,
where the systematic error was estimated by varying the binning and range
used to scale the correlation function for  $0.5<|\Delta \eta|<1.4$.
This width is consistent with earlier observations of 
jet characteristics in $pp$ collisions at slightly lower \cite{SFM} and higher \cite{UA1} energies. The HIJING event generator \cite{hijing}, 
where hard scattering and fragmentation dominate particle production at 
these transverse momenta, predicts $\sigma=0.20\pm0.01$ for the same kinematic cuts.. Integrating the signal we observe that 
$4.9 \pm 1.7$(stat.)$\pm0.4$(sys.) \%  of charged particles 
with $p_T$ of 4-6 GeV/c have an associated charged particle with $p_T>2$ GeV/c.
The systematic error is dominated by the uncertainty in the absolute efficiency
determined via the embedding procedure.
The contributions of resonance decays and photon conversions were studied and 
found to be insignificant.

In summary, the measurements of azimuthal ani\-sotro\-py $v_2$ of charged
particles with $p_T$ of 3-6 GeV/c 
reveal a saturation pattern of $v_2$
with values that decrease systematically with increasing centrality.
This contradicts non-dissipative hydrodynamics which predicts a
monotonically increasing
$v_2$ with increasing $p_T$, but the data 
may be consistent with dissipative dynamics with finite
parton energy loss.
In addition, a comparison of the two-particle azimuthal correlation 
functions for
particles with $|\Delta\eta|<0.5$ and $|\Delta\eta|>0.5$ 
suggests the existence of a
short-range correlated
component at high $p_T$ in addition to underlying global elliptic
flow. This
may be the first direct evidence at RHIC for hard scattering and parton
fragmentation.
The data provide important constraints on the
theoretical interpretations of the mechanism of high $p_T$ particle
production in ultra-relativistic heavy-ion collisions.

\begin{acknowledgments}

We wish to thank the RHIC Operations Group and the RHIC Computing Facility
at Brookhaven National Laboratory, and the National Energy Research 
Scientific Computing Center at Lawrence Berkeley National Laboratory
for their support. This work was supported by the Division of Nuclear 
Physics and the Division of High Energy Physics of the Office of Science of 
the U.S. Department of Energy, the United States National Science Foundation,
the Bundesministerium f\"ur Bildung und Forschung of Germany,
the Institut National de la Physique Nucl\'eaire et de la Physique 
des Particules of France, the United Kingdom Engineering and Physical 
Sciences Research Council, Funda\c c\~ao de Amparo \`a Pesquisa do Estado de S\~ao Paulo, Brazil, the Russian Ministry of Science and Technology and the
Ministry of Education of China and the National Science Foundation of China.

\end{acknowledgments}

\bibliographystyle{unsrt}

\begin{thebibliography}{99}

\bibitem{v2charged} 
STAR Collaboration, K.~H.~Ackermann \etal, \PRL {\bf 86}, 402 (2001).
\bibitem{olli} J.-Y.~Ollitrault, \PRD {\bf 46}, 229 (1992).
\bibitem{sorge} H.~Sorge, \PRL {\bf 78}, 2309 (1997).
\bibitem{energyloss} 
M.~Gyulassy and M.~Plumer, Phys. Lett. B {\bf 243}, 432 (1990); 
X.-N.~Wang and M.~Gyulassy, \PRL {\bf 68}, 1480 (1992).
\bibitem{energyloss2}
M.~Gyulassy and X.-N.~Wang,
Nucl. Phys. B {\bf 420}, 583 (1994); R.~Baier, 
Y.~L.~Dokshitzer, S.~Peigne and D.~Schiff, Phys. Lett. B {\bf 345}, 277 (1995).
\bibitem{phenix} 
PHENIX Collaboration, K.~Adcox \etal, \PRL {\bf 88}, 022301 (2002).
\bibitem{StarHighpt} STAR Collaboration, C.~Adler \etal, \PRL {\bf 89}, 202301 (2002). 
\bibitem{wang}X.-N.~Wang, \PRC {\bf 63}, 054902 (2001).
\bibitem{glv}M.~Gyulassy, I.~Vitev and X.-N.~Wang, \PRL {\bf 86}, 2537 (2001).
\bibitem{voloshin} 
S.~Voloshin and Y.~Zhang, Z. Phys. C {\bf 70}, 665 (1996); 
A.~M.~Poskanzer and S.~A.~Voloshin, \PRC {\bf 58}, 1671 (1998).
\bibitem{v2identified} 
STAR Collaboration, C.~Adler \etal, \PRL {\bf 87}, 182301 (2001).
\bibitem{nimTPC} M.~Anderson \etal, Nucl. Instr. Meth. A 
(2002), RHIC Special Volume, in press. 
\bibitem{UA1} 
G.~Arnison \etal, \PLB {\bf 118}, 173 (1982).
\bibitem{4part} 
N.~Borghini, P.~M.~Dinh, and J.-Y.~Ollitrault, \PRC {\bf 64}, 054901 (2001).
\bibitem{flowprc} 
STAR Collaboration, C.~Adler \etal, \PRC {\bf 66}, 034904 (2002). 
\bibitem{hydro} 
P.~Huovinen, P.~F.~Kolb, U.~W.~Heinz, P.~V.~Ruuskanen, 
and S.~A.~Voloshin, Phys. Lett. B {\bf 503}, 58 (2001).
\bibitem{glvp}
M.~Gyulassy, I.~Vitev, X.-N.~Wang, 
and P.~Huovinen, Phys. Lett. B {\bf 526}, 301 (2002).
\bibitem{molnar}
D.~Molnar and M.~Gyulassy, Nucl. Phys. A {\bf 697}, 495 (2002).
\bibitem{SFM} 
A.~Breakstone \etal, Z. Phys. C {\bf 23}, 1 (1984).
\bibitem{hijing} 
X.-N.~Wang, \PRD  {\bf 46}, R1900 (1992); \PRD {\bf 47}, 2754 (1993).

\end{thebibliography}

\end{document}